# Dynamic disruption index across citation and cited references windows: Recommendations for thresholds in research evaluation

Hongkan Chen[1], Lutz Bornmann[2*], Yi Bu[1,3,4*]

1 Department of Information Management

Peking University

5 Yiheyuan Road, Haidian District

100871 Beijing, China.

2 Science Policy and Strategy Department

Administrative Headquarters of the Max Planck Society

Hofgartenstr. 8,

80539 Munich, Germany.

3 Center for Informationalization and Information Management

Peking University

5 Yiheyuan Road, Haidian District

100871 Beijing, China.

4 Peking University Chongqing Research Institute of Big Data

Building 10, Science Valley, High-Tech Zone

401332 Chongqing, China.

* Corresponding author's email: chenhongkan@pku.edu.cn

# Abstract

The temporal dimension of citation accumulation poses fundamental challenges for quantitative research evaluations, particularly in assessing disruptive and consolidating research through the disruption index (D). While prior studies emphasize minimum citation windows (mostly 3–5 years) for reliable citation impact measurements, the time-sensitive nature of D—which quantifies a paper's capacity to eclipse prior knowledge—remains underexplored. This study addresses two critical gaps: (1) determining the temporal thresholds required for publications to meet citation/reference prerequisites, and (2) identifying "optimal" citation windows that balance early predictability and longitudinal validity. By analyzing millions of publications across four fields with varying citation dynamics, we employ some metrics to track D stabilization patterns. Key findings reveal that a 10-year window achieves >80% agreement with final D classifications, while shorter windows (3 years) exhibit instability. Publications with ≥30 references stabilize 1–3 years faster, and extreme cases (top/bottom 5% D values) become identifiable within 5 years—enabling early detection of 60–80% of highly disruptive and consolidating works. The findings offer significant implications for scholarly evaluation and science policy, emphasizing the need for careful consideration of citation window length in research assessment (based on D).

# Introduction

The number of citations a certain publication has received is a function of time: the longer the time between the dates of collecting citations and publication, the more citations can be expected. In the natural and life sciences, the number of citations tends to reach its highest point on average during the third or fourth year following publication (Van Raan, 2019). Since decades, research in bibliometrics has dealt with the question of an appropriate length of citation windows in citation analyses (e.g., Blanckenberg & Swart, 2018; Wang, 2013). Most of the literature has focused on the minimum citation window that should be considered in these analyses. The recommendations can be summarized as follows: longer citation windows lead to more reliable impact assessments (Wang, 2013); the citation window should have a minimum of 3 to 5 years.

Wu et al. (2019) introduce the disruption index (D) that proposes to measure disruptiveness and consolidating nature of papers based on citation and cited references data. As an overview of research on D (Leibel & Bornmann, 2024b) mentions that D suffers from inconsistency, time-sensitive biases, and several data-induced biases, we wonder how the D value of a publication changes with the extension of the citation window in a given dataset. We aim to better understand the impact of inconsistency, time-sensitive biases, and several data-induced biases on D, particularly by addressing potential discrepancies in the sign of disruption values: Are there discrepancies whether a publication is denoted as disruptive or consolidating, between the initial years and the present (Bornmann & Tekles, 2019)? In this study, we aim to quantify how the citation window influences D values and to assist researchers in establishing meaningful thresholds for calculating D.

Let us imagine what factors a scientist needs to consider when applying D. When selecting publications, aside from disregarding factors such as discipline, journal, and authors, another aspect is to choose publications with a certain number of cited references and citations collected across a certain number of years. One may expect that the number of references in a publication is fixed and is only influenced by data-induced biases. However, discrepancies may arise in datasets such as Web of Science (Clarivate) and Scopus (Elsevier) due to missing publication entities or citation linkages. Citation counts for a publication tend to increase non-strictly over time, making the length of the citation window a critical consideration. This window not only determines metrics like citation counts that measure a publication's impact but also influence the D value itself, as D is time-sensitive. Since the citations and cited references of a publication are crucial elements in calculating D values, thresholds for counting time and minimum numbers are essential information. The primary goal of this study is to better assist researchers in determining meaningful thresholds.

We focus on two research questions in this study. First, considering that citation counts are closely tied to the citation window, we investigate the time span required for publications to reach a particular citation threshold for inclusion in certain D analyses. This analysis serves as a foundation for setting thresholds on the number of citations (or cited references) for calculating D values. The overview of Leibel and Bornmann (2024b) of research on D shows that previous studies have used various thresholds. The authors recommend "to calculate disruption scores only for publications with at least a certain number (e.g., ten) of citations and cited references" (p. 633). However, it remained an open question which thresholds exactly should be used. Second, after simply restricting the number of references and citations, we explore how D values change over time, thereby determining the optimal citation window length. Researchers have used various citation windows in their use of D – see the overview in Leibel & Bornmann (2024a). Bornmann and Tekles (2019) demonstrate—based on only four papers—that D values depend on the citation window. Therefore, we examine in this study in a first step the differences (relationships) between the D values calculated using different time windows and the final D value based on a comprehensive dataset. In a second step, we investigate when D values stabilize, which helps to better determine the optimal citation window length.

# Data

The dataset utilized in this study is derived from OpenAlex (snapshot from 27/02/2024), a comprehensive and openly accessible knowledge graph that aggregates metadata on scholarly works across a wide range of disciplines. OpenAlex organizes its data into four main domains—social sciences, natural sciences, life sciences, and health sciences—each encompassing multiple fields of study. To ensure a balanced representation across these domains, we select one representative field (labelled by OpenAlex) from each: economics, econometrics, and finance (social sciences), immunology and microbiology (life sciences), physics and astronomy (natural sciences), and nursing (health sciences). We use the selected fields to investigate whether there are field-dependencies in our results. Descriptive statistics of the used dataset in this study can be found in Table S1.

For each field, we focus on the articles published in journals or conference proceedings. Figure 1 illustrates the cumulative distributions for the analyzed publications, depicting the distribution patterns of both cited reference counts and citation counts. The figure demonstrates that most publications exhibit low reference and citation counts. While it is usual for many publications to have no or a few citations, a significant number of publications still lack references, even after restricting to journals and conference proceedings. This raises serious concerns and contradicts established academic norms, as complete absence of references is exceptionally rare in practice. The discrepancy indicates substantial gaps in database coverage, where not all cited references are covered as source items. Such data incompleteness introduces systematic biases, particularly affecting the reliability of D (Leibel & Bornmann, 2024b). Specifically, missing reference data may lead to inaccurate identification of disruptive and consolidating patterns, as the index relies heavily on comprehensive citation networks. This limitation underscores the need for caution when interpreting results derived from bibliometric datasets, especially in research using D.

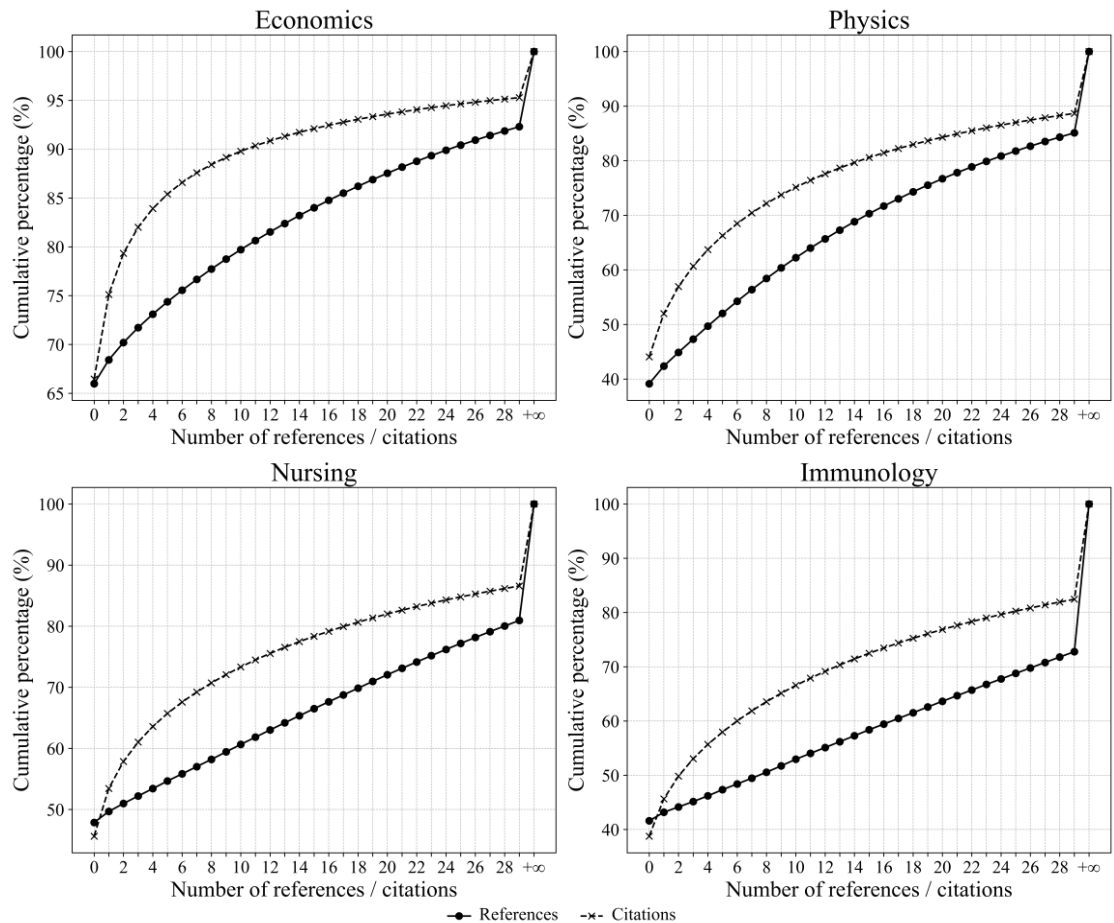

Figure 1. Reference and citation count distributions across publications published in journals or conference proceedings.

Besides, when applying a threshold of 5 for both cited reference counts and citation counts, as discussed in the second research question in economics, over 70% of publications have fewer than 5 references, and more than 80% have fewer than 5 citations. This indicates that publications qualifying for D analyses tend to be more impactful (in terms of citations) compared to the majority and only those with relatively high citation counts meet the criteria for evaluating their disruptive and consolidating nature.

# Disruption

D originates from analyses of technological innovation, where Funk & Owen-Smith (2017) conceptualize it as a metric to quantify the magnitude of structural change introduced by patents. Unlike conventional novelty metrics (e.g., Foster et al., 2015; Lee et al., 2015; Uzzi et al., 2013), which primarily assess incremental originality, D explicitly captures the "displacement" effect of an invention on prior knowledge networks. This "displacement" framework was transferred from patents to scientific publications by L. Wu et al. (2019), who demonstrated its applicability to bibliometric contexts. The adaptation leverages the structural parallels between technological patents and scientific publications, particularly their shared reliance on citation dynamics to trace knowledge flows. For a focal publication (FP), D is defined as:

$$D = \frac{N_F - N_B}{N_F + N_B + N_R}$$

where $N_F$ denotes the number of publications citing only the FP, excluding any references cited by the FP; $N_B$ represents the number of publications citing both the FP and at least one reference from the FP; $N_R$ denotes the number of publications that appear after the FP and cite at least one reference of the FP but not the FP itself. This formulation operationalizes disruption as the difference between "exclusive" citations ($N_F$), signaling replacement of prior knowledge, and "overlapping" citations ($N_B$), signaling continuity. $N_R$ serves as a normalization factor, reflecting the baseline influence of the FP's cited references set. D values are between 1 (maximally disruptive and -1 (maximally consolidating).

In the years following its introduction, D has attracted significant attention from the scientometrics and quantitative science studies communities (and beyond). Researchers have critically engaged with this metric, proposing various modifications to address its identified limitations. An overview of the D research has been published by Leibel & Bornmann (2024). This study focuses on exploring the temporal evolution of D values to provide practical guidance for setting citation windows in empirical applications. Recent studies have highlighted critical limitations in the robustness of D. Ruan et al. (2021) demonstrate that D exhibits significant sensitivity to the number of references in a publication. Holst et al. (2024) further reveal its vulnerability to data entries with zero cited references, which can distort derived results. To mitigate these issues, we adopt the methodological recommendation by Leibel and Bornmann (2024b). The authors propose to filter publications based on minimum thresholds for both forward citations and references. For the second research question in this study, we operationalize the approach by setting the threshold to 5 for both sides when calculating D values.

We distinguish between two sights on disruptiveness and consolidating nature of papers for the second research question. The first sight is the "true" D level of a paper. We assume that this is a constant paper characteristic which does not change over time. To operationalize this concept, we utilize the sign of the D value (i.e., whether a paper is disruptive or consolidating) and its inclusion in top-ranked disruptive or consolidating publications as characteristics. The second sight is the measurement of this constant characteristic: In our study, we try to determine when the "true" disruption or consolidating level can be measured (e.g., with a 3- or 5-year window). For operational clarity, we define the final D values—calculated without citation window constraints, using all available citation data in the dataset—as the baseline ground truth. Methodological details are provided in the following sections.

# Methods

## Time to stabilization of impact metrics

To quantify the time required for the selected publications to qualify for disruption analysis (in terms of reliability), we formally define two measurement functions. First, we establish the indicator *citation time lag* (CTL) to capture citation latency patterns:

$$CTL(P, x) \coloneqq \{t_p^x | \forall\, p \in P\}$$

$$t_p^x \coloneqq citation\_year(p, x) - publication\_year(p)$$

where $P$ denotes the target publication set. $publication\_year(p)$ indicates the publication year of $p$. $citation\_year(p, x)$ is the calendar year when $p$ received its $x$th citation. Thus, $t_p^x$ represents how long $p$ received its $x$th citation.

Although we set thresholds for cited references and forward citations (Leibel & Bornmann, 2024b), we only set a parameter for the number of forward citations while the reference counts is a constant variable when we consider citation windows in this part. Therefore, we construct *aggregated citation time lag* to indicate the most representative value of the time lag from CTLs of all publications that meet the specified criteria:

$$Agg\_CTL(f, x, y) \coloneqq f(CTL(P, y))$$

$$P \coloneqq \{\forall p\ |\ \#references(p) \geq x, \#citations(p) \geq y\}$$

where $x$ and $y$ are two thresholds of filters for the number of cited references and forward citations, respectively. $f(*)$ is an aggregating method to transform a list of numbers (e.g., $CTL(P, x)$) into one value. $f(*)$ allows flexible applications of different analytical operations (e.g., statistical means or different quantiles) on derived time intervals. $P$ represents all publications satisfied with $x$ references and $y$ citations when the citation window is unlimited. $\#citations(p)$ represents citation counts of $p$ in the dataset while $\#references(p)$ represents cited reference counts.

## Time to stabilization of D values

D values might change with new citations. Therefore, we define stabilization as the year when the sign of D values no longer changed until the end year – defined by the used dataset (i.e., the publication maintains disruptive or consolidating since then). We define *stabilized year of disruption* (SYD) for each publication:

$$SYD(p) := min\{t \in N | \forall k \geq t, sign(d_p^k) = sign(d_p^{final})\}$$

where $d_p^t$ represents the D value of publication $p$ after $t$ years and $d_p^{final}$ equals to $d_p^{\infty}$ for D values of publications without any citation window limitation (i.e., final D values). In this study, only publications with at least 5 cited references and citations have D values; otherwise $d_p^t$ is null. Hence, for a group of publications, we define *selected stabilized year of disruption* with thresholds of citation counts as:

$$Selected\_SYD(P, x) := \{dt_p^x | \forall\, p \in P\}$$

$$dt_p^x = max(t_p^x, DSY(p))$$

where $dt_p^x$ represents the time when publication $p$ both achieves a stable D value and accumulates at least $x$ citations. We also define *aggregated selected stabilized year of disruption* to measure the time required for the publications' D value to get stabilized:

$$Agg\_selected\_SYD(f, x, y) := f(Selected\_SYD(P, y))$$

$$P := \{\forall p \mid \#references(p) \geq x, \#citations(p) \geq y\}$$

## Temporal correlation of D values

What is the degree of similarity between D values derived from different time windows and the "final" D values? We define final D values as those values that are computed without citation-window constraints in this study. We employ four correlation and concordance coefficients, respectively: Pearson correlation coefficient (Pearson), Spearman's rank coefficient (Spearman), Kendall rank coefficient (Kendall), and Lin's concordance correlation coefficient (Lin's CCC). Each coefficient has its own strength and limitations: Pearson measures linear relationships, suitable for normally distributed data. Spearman assesses monotonic relationships, based on rank order. Kendall evaluates rank-based associations, robust to non-normal and non-monotonic data. Lin's CCC quantifies agreement by assessing both precision and accuracy relative to the identity line.

Pearson and Spearman are standard coefficients usually used in studies measuring correlations. However, Kendall is chosen as the primary metric in this study because D values are neither normally distributed (invalidating Pearson) nor consistently monotonic (limiting Spearman). Kendall is non-parametric and robust to outliers, making it ideal for assessing rank-based similarity across time windows. Lin's CCC is additionally considered to evaluate concordance, as high correlation alone does not guarantee agreement. For example, systematic over- or under-estimation of D values across citation windows could yield high correlations but poor concordance. Lin's CCC addresses this by measuring both precision (deviation from the best-fit line) and accuracy (deviation from the identity line). High coefficients refer to D values that are not only correlated but also consistent. Our multi-method approach using various coefficients ensures a rigorous and nuanced analysis of the similarity between D values at different time windows and their final values.

To calculate the similarity between D values at a specific time window $t$ ($d_p^t$) and the final D values ($d_p^{final}$), we use Kendall ($\tau$):

$$\tau(t) = \frac{C - D}{\frac{1}{2}n(n-1)}$$

where $C$ is the number of concordant pairs ($d_p^t$ and $d_p^{final}$ agree in rank order), $D$ is the number of discordant pairs ($d_p^t$ and $d_p^{final}$ disagree in rank order), and $n$ is the total number of pairs.

## Reliability of early D assessments

To evaluate the reliability of early measured D values, we analyze the temporal dynamics of publications with high D values through a multi-metric approach. Specifically, we quantify the proportion of highly disruptive publications and their persistence over time, providing insights into the stability and predictive validity of early disruptiveness signals. The methodological framework is implemented by the following steps:

1. Yearly highly disruptive publication threshold

The *yearly highly disruptive publication threshold* (YHT) of a paper set is determined

as the 90th percentile of D values for all publications in this set $t$ years after their publication:

$$YHT_t := \text{Quantile}_{0.9}\big(\{d_p^t \mid p \in P\}\big)$$

where $\text{Quantile}_{0.9}(X)$ represents the 90th value in $X$.

2. Final highly disruptive publication threshold

The *final highly disruptive publication threshold* (FHT) of a paper set is calculated as the 90th percentile of the final D values across all publications:

$$FHT := \text{Quantile}_{0.9}\big(\{d_p^{\text{final}} \mid p \in P\}\big)$$

3. Yearly highly disruptive publication ratio

For each year $t$ after publication, we calculate the proportion of highly disruptive publications (top 10%) relative to all publications with available D values at $t$ years after publication. The *yearly highly disruptive publication ratio* is defined as:

$$Yearly\_highly\_disruptive\_publication\_ratio_t := \frac{\big|\{p \mid d_p^t \geq YHT_t\}\big|}{\big|\{p \mid d_p^t \text{ is not null}\}\big|}$$

4. Overlapping ratio

We evaluate the proportion of yearly publications with high disruptiveness that maintain their high status in the final disruptiveness value. This *overlapping ratio* is computed as:

$$Overlapping\_ratio_t := \frac{\big|\{p \mid d_p^t \geq YHT_t \wedge d_p^{\text{final}} \geq FHT\}\big|}{\big|\{p \mid d_p^t \geq YHT_t\}\big|}$$

5. Early identified ratio

We assess the proportion of publications with final high disruptiveness that are classified as highly disruptive in year $t$. This *early identified ratio* is defined as:

$$Early\_identified\_ratio_t := \frac{\big|\{p \mid d_p^{\text{final}} \geq FHT \wedge d_p^t \geq YHT_t\}\big|}{\big|\{p \mid d_p^{\text{final}} \geq FHT\}\big|}$$

The yearly highly disruptive publication ratio reflects the prevalence of highly disruptive publications at $t$ years after publication. The overlapping ratio indicates the likelihood that yearly publications with high disruptiveness persist as highly disruptive in the long term. The early identified ratio measures the ability to predict, at an early stage, which publications will ultimately have high disruptiveness values. These metrics enable a robust evaluation of the temporal dynamics of highly disruptive publications, offering insights into their early identification and long-term stability.

## Temporal consistency of D values

Moving from the most disruptive publication sample to the entire sample, we attempt to describe whether the disruptive and consolidating nature of publications can be reliably measured. To analyze the consistency of D values over time, we calculate three ratios. The *available disruptiveness data ratio* serves as a measure of data coverage, indicating the proportion of publications for which D values are available at $t$ years after publication. Note that only publications with at least 5 cited references and 5 citations are considered here with their D values. The *consistent ratio* reflects the overall consistency of D value signs across the entire dataset, capturing how often the research type (disruptive or consolidating) at a given year after publication aligns with the final type. Meanwhile, the *yearly consistent ratio* specifically evaluates the reliability of types for publications with available data at $t$ years after publication, providing insights into the accuracy of early assessments.

The proposed metrics offer a comprehensive understanding of the temporal consistency of D values, shedding light on the stability and predictive power of D over time. The metrics not only investigate the reliability of early measurements but also underscores the dynamic evolution of D values. $N$ in the formulas is the total number of publications:

$$Available\_disruptiveness\_data\_ratio_t := \frac{|\{p \mid d_p^t \text{ is not null}\}|}{N}$$

$$Consistent\_ratio_t := \frac{|\{p \mid \text{sign}(d_p^t) = \text{sign}(d_p^{\text{final}})\}|}{N}$$

$$Yearly\_consistent\_ratio_t := \frac{|\{p \mid \text{sign}(d_p^t) = \text{sign}(d_p^{\text{final}})\}|}{|\{p \mid d_p^t \text{ is not null}\}|}$$

## Volatility and stability of D values

To analyze the stability of D values and their rankings, we introduce three metrics: the *percentage change in D values*, the *absolute change in D values*, and the *change of normalized ranking*.

The *percentage change in D values* measures the relative change in D from one year to the next:

$$P_p(t) := \frac{|d_p^{t+1} - d_p^t|}{|d_p^t|}$$

This metric highlights the volatility of D values over time, with higher values indicating greater instability.

The *absolute change in D values* quantifies the magnitude of change without considering the direction:

$$A_p(t) := |d_p^{t+1} - d_p^t|$$

This metric provides insights into the scale of D value fluctuations, which can be particularly useful for identifying significant shifts in a publication's disruptiveness or consolidating nature.

To address the issue of cross-time comparability, we normalize the D value of publications within each year, assigning values from 0 to 100 based on their D value rankings. The *change of normalized ranking* is then computed as:

$$R_p(t) := |\text{Rank}_p(t+1) - \text{Rank}_p(t)|$$

where $\text{Rank}_p(t)$ represents the normalized rank of publication $p$ within a group of publications, all published in the same year as $p$. The metric captures the relative movement of publications within their annual rankings, offering a clearer picture of their D value trends over time.

Percentage change and absolute change metrics reveal the degree of instability in D values, with higher values indicating greater fluctuations. The change of normalized ranking provides a time-agnostic view of how publications' disruptiveness or consolidating nature evolves relative to their peers. These metrics collectively offer a robust framework for understanding the stability and evolution of D values and rankings, enabling the assessment of long-term impact and consistency of D.

## Transition pattern between disruptiveness and consolidation

To investigate whether publications undergo significant shifts between disruptiveness ($D > 0$) and consolidation ($D \leq 0$) we adopt a categorical encoding approach. We categorize publications into the following groups based on D:

- *Stable*: Publications that remain consistently disruptive or consolidating from the first D year is calculable (i.e., the year when the publications receive their fifth citations).
- *Disruptive_consolidating*: Publications that are disruptive in their first year (i.e., the first year for which a D value could be calculated) and then remain consolidating thereafter.
- *Consolidating_disruptive* similarly): Publications that are consolidating in their first year and then remain disruptive thereafter.
- *>1_disruptive_consolidating*: Publications that are disruptive for several years, after which they transition to consolidating.
- *>1_consolidating_disruptive*: Publications that are consolidating for several years, after which they transition to disruptive.
- *Highly_unstable*: Publications that exhibit at least two transitions, where each transition signifies a shift between disruptive and consolidating.

We count the number of publications with different transition frequencies for publications of *Highly_unstable* type. This methodology allows us to quantify the stability and variability of categories over time, providing insights into the dynamic nature of disruption and consolidation.

D values around zero may not reveal much about the disruptive or consolidating nature of papers. They may be random variations. Thus, we also analyze data where only publications with D values in the top 10% are assigned disruptive, and all others are assigned consolidating. We do the same for consolidating publications (bottom 10% D values).

# Results

## Length of time to achieve minimum citation counts across cited reference and citation thresholds

Figure 2 presents a heatmap of $Agg\_CTL(average,*,*)$, illustrating the average time required for publications to achieve specific citation thresholds under varying conditions of cited references and citations. For instance, for economics publications with at least 5 cited references and 5 citations, the average time to achieve the fifth citation is 4.08 years; with a fixed number of cited references, the average time increases as the minimum citation counts grow. This ranges from 2.10 years for the first citation to 8.54 years for the 30th citation when restricting the minimum cited references as one. It ranges from 1.14 to 5.59 years when restricting the minimum cited references to 30. The results also show that publications with a higher number of cited references tend to be cited earlier. For instance, publications with at least 30 cited references achieve their first citation nearly one year earlier than those with at least one reference (Bornmann & Daniel, 2008; Tahamtan & Bornmann, 2019). For the 30th citation, the year difference grows to 2.95 years. The results in Figure 2 reveal that they are robust in different fields.

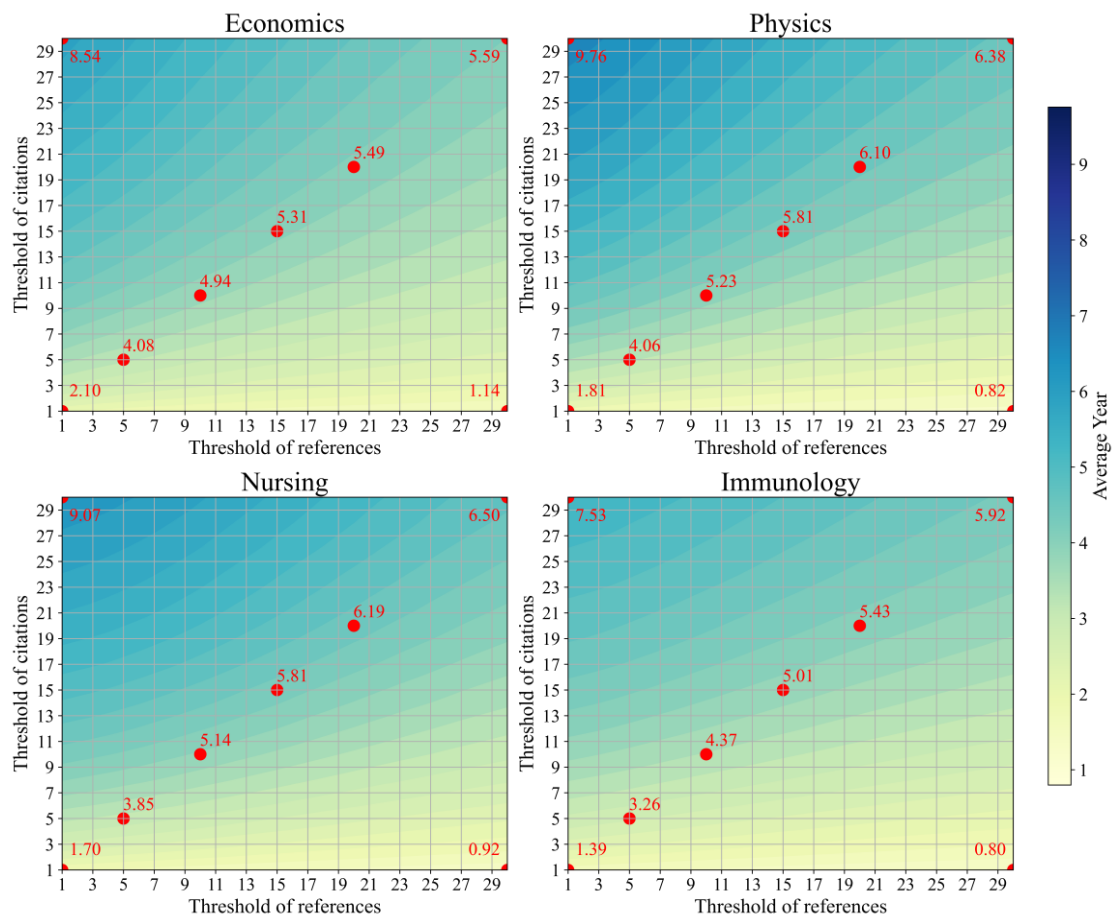

Figure 2. Average time to achieve minimum citation counts across cited reference and citation thresholds. Each heatmap illustrates the average time (in years) required for publications to achieve a specified minimum citation count, as a function of the minimum number of references required for inclusion in the analysis. The x-axis represents the minimum number of references, and the y-axis represents the minimum number of citations.

We analyze the time required for different proportions of publications to achieve the required citation counts under various thresholds combinations, as shown in Figure 3. If we focus on the 50th and 80th quantile on the x-axis, we find that the median year is consistently lower than the average year, while the 80th percentile year is higher. The results in Figure 3 demonstrate that a citation window of 3 years is insufficient to encompass the impact of the vast majority (80%) of publications. The 10-year citation window seems to be sufficient instead.

Table S2 exemplarily provides the number of publications in different years across different thresholds combinations in economics.

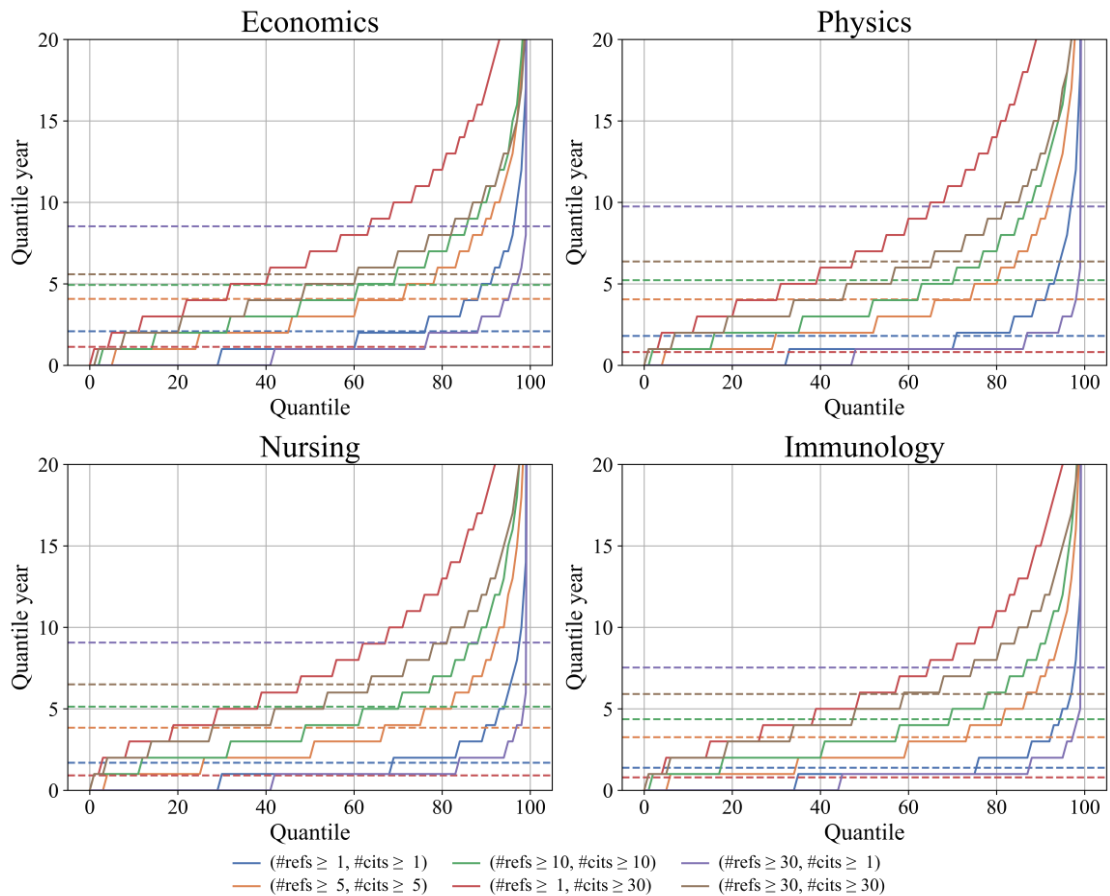

Figure 3. Quantile analysis of time to achieve minimum citation counts. The figure presents the distribution of time (in years) required for publications to achieve specified minimum citation counts, focusing on selected combinations of cited reference and citation thresholds. The x-axis represents publication quantiles, and the y-axis the corresponding time values. The horizontal line represents the average time (the same as Figure 2) of the corresponding color combination.

## Disruption with multiple time windows

In the analyses of D values, we restrict the publications to those with at least 5 cited references and 5 citations. Figure S1 illustrates the distribution of final D values across publications from different years. The results do not show significant differences in the distributions across different publication year restrictions. Many D values are near to zero. As Figure 4 shows, in the 3-year citation window, all correlation coefficients are below 0.8, which indicates that the 3-year citation window may not be suitable to reliably measure disruption and consolidation. The results show that the longer the citation window, the more reliable D is.

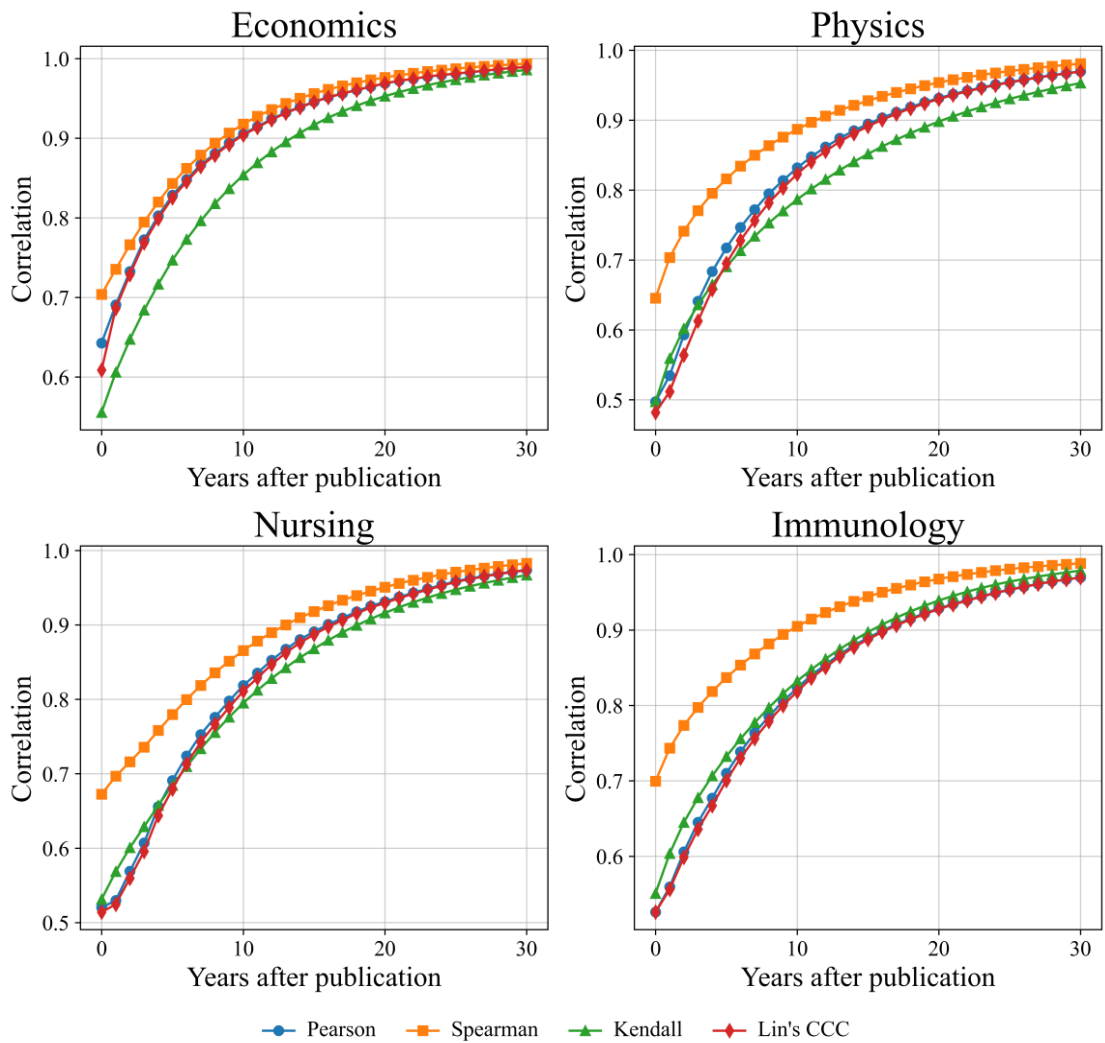

Figure 4. Field-specific correlation and concordance coefficients for the relationship of D measured with different time windows and final D.

Figure 5 groups publications based on their final D values (positive or negative) and plots the changes in Kendall over time windows for 5 deciles. A clear overall trend emerges: the correlation between disruption values and final disruption values monotonically increases with longer citation windows across all publication categories. It also demonstrates that consolidating publications (blue curves) consistently exhibit higher correlation coefficients than disruptive ones (red curves) across all time windows, with the most extreme groups—the highest consolidating (D<0) and most disruptive (D>0) publications—showing the strongest early correlations within their respective categories. This dual hierarchy reveals that consolidating patterns are inherently more predictable in short-term citation data, while the earliest identification accuracy concentrates on publications with polarized D values (either strongly consolidating or disruptive). These results validate the feasibility of using limited time windows for highly confident detection of extreme consolidating/disruptive publications, though full categorization requires extended observation to resolve ambiguous mid-range D values.

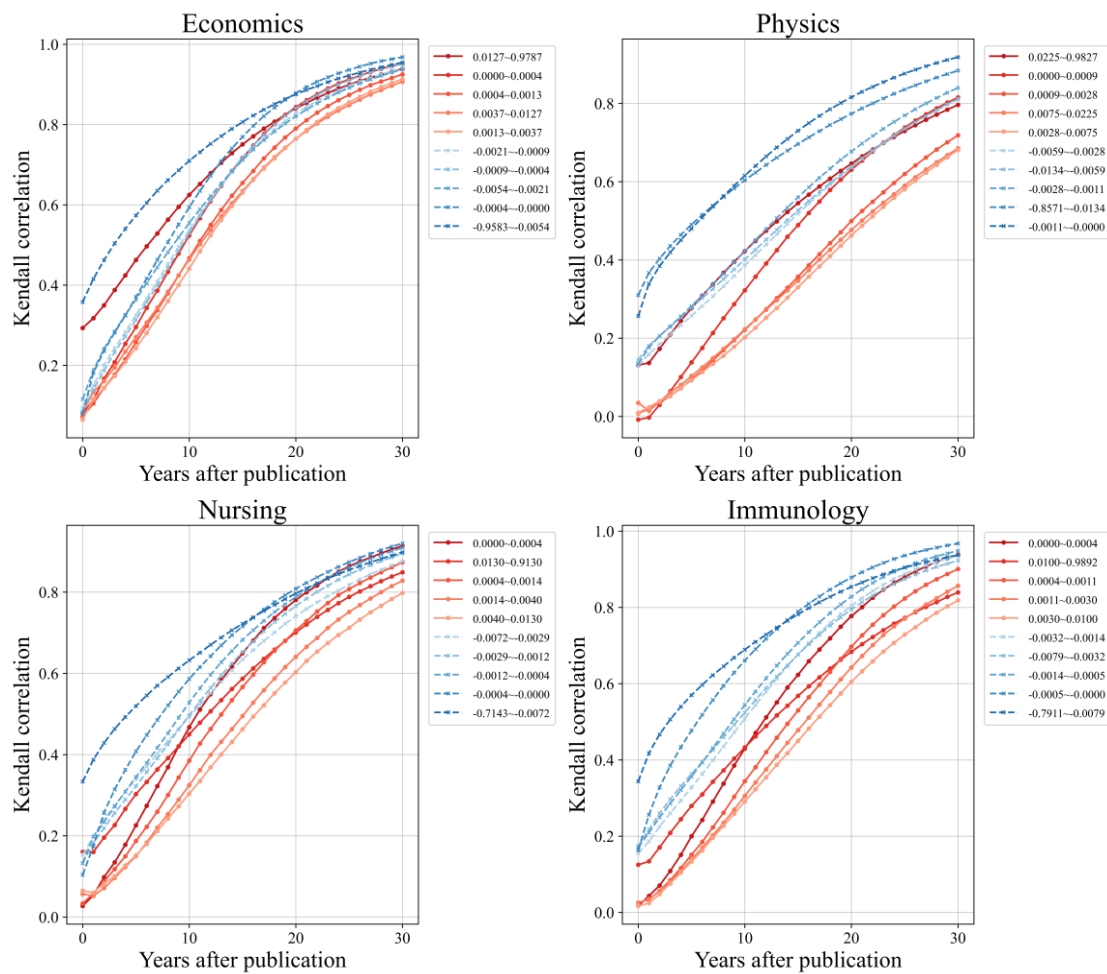

Figure 5. Field-specific temporal evolution of Kendall for publications grouped by final D values.

## Potential of early identifying highly disruptive papers

Due to the prevalence of publications with consistently low and similar citation counts, more than 10% of publications may reach the top 10% within the first two years, as citations take time to accumulate (Wang, 2013). In other words, influential publications cannot be clearly distinguished in their early years. However, the calculation of D involves a threshold on citation counts, which ensures that disruptiveness is distinguishable from the outset. This is supported by the yearly highly disruptive publication ratio, which stabilizes at around 10% consistently, as shown in Figure 6.

In economics, the overlapping ratio reveals that nearly 70% of highly disruptive publications maintain their status by the zeroth year, increasing to about 75% by the fifth year and surpassing 80% by the tenth year. From the perspective of final highly disruptive publications, the early identification ratio grows from zero to nearly 100%, exhibiting a convex shape. Specifically, the early identification ratio shows that more than 60% of highly disruptive publications are identified by the fifth year, and nearly 80% by the tenth year. In the other fields, the overlapping ratio closely aligns with the early identification ratio at economics. The highly disruptive publications in the final dataset are identified at a rate of 50% by the fifth year and approximately 70% by the tenth year.

Additionally, while the yearly highly disruptive publication threshold in economics shows a high value in the zeroth year, it drops rapidly below the final highly disruptive publication threshold in the first year and then gradually rises, approaching it by the third year. This trend suggests two key insights for economics: (1) The disruptiveness threshold exhibits significant fluctuations in the early years, making short citation windows less reliable as a restriction. (2) As the time window expands, the threshold for highly disruptive publications increases, indicating a growing entry barrier for such publications. This phenomenon may be attributed to the rising disruptiveness of highly disruptive publications over time. However, the results from nursing and immunology exhibit the growing trend and little fluctuations of yearly highly disruptive publication thresholds in the first two years.

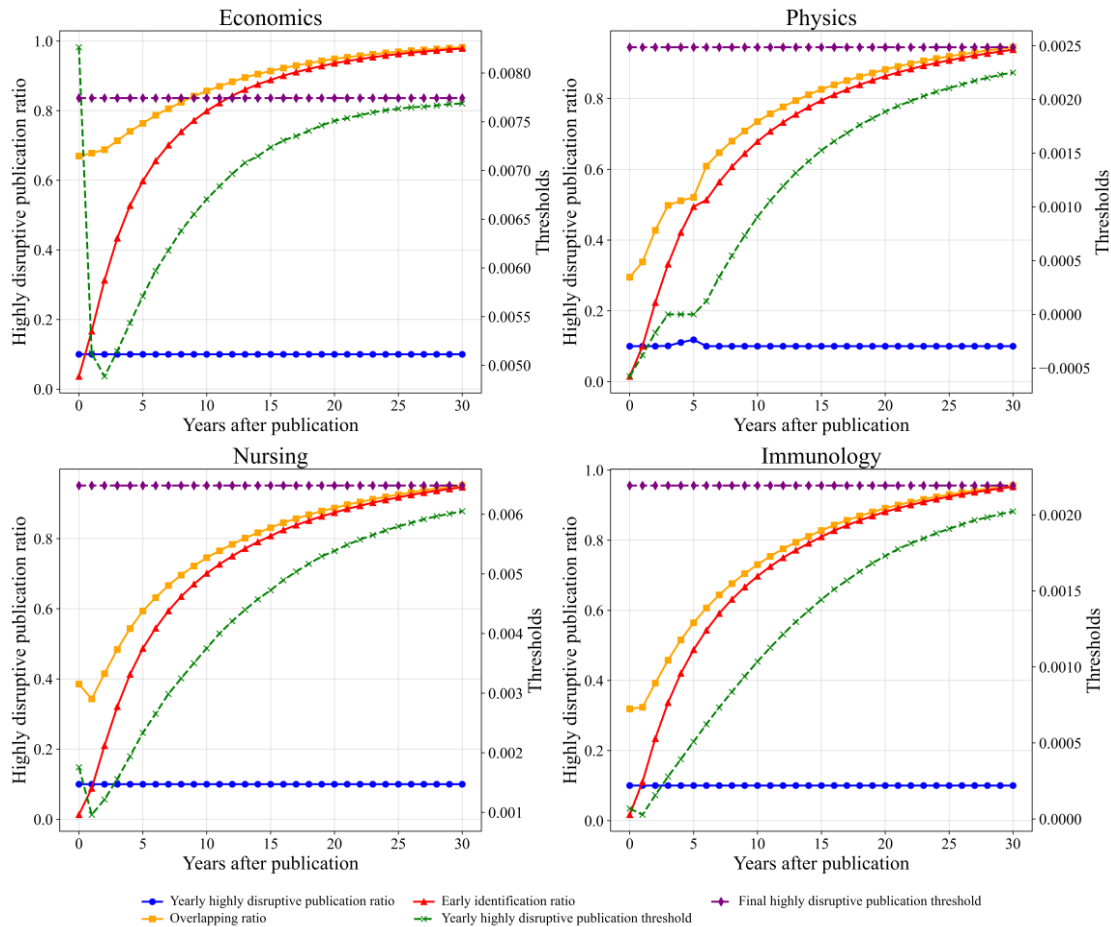

Figure 6. Temporal characteristics of highly disruptive publications.

Figure 7 investigates whether the fluctuations of the yearly highly disruptive publication threshold in early years represent a regular phenomenon. The results reveal that the most pronounced trends occur among highly disruptive publications (Rank 0.1% and 0.5%) and highly consolidating publications (99.9% and 99.5%). After approximately 5 years, the values across all percentiles stabilize and remain relatively constant. Overall, however, the D values in the early years approach zero, suggesting that early estimates of disruption are relatively inflated in absolute terms. This phenomenon may be attributed to the increasing $N_R$ in the calculation of disruption, highlighting the weakness of D—its inconsistency. Furthermore, the values for all ranks, whether disruptive or consolidating, exhibit growth over time. This confirms that the increase in D values is a widespread trend that can be observed across all publications.

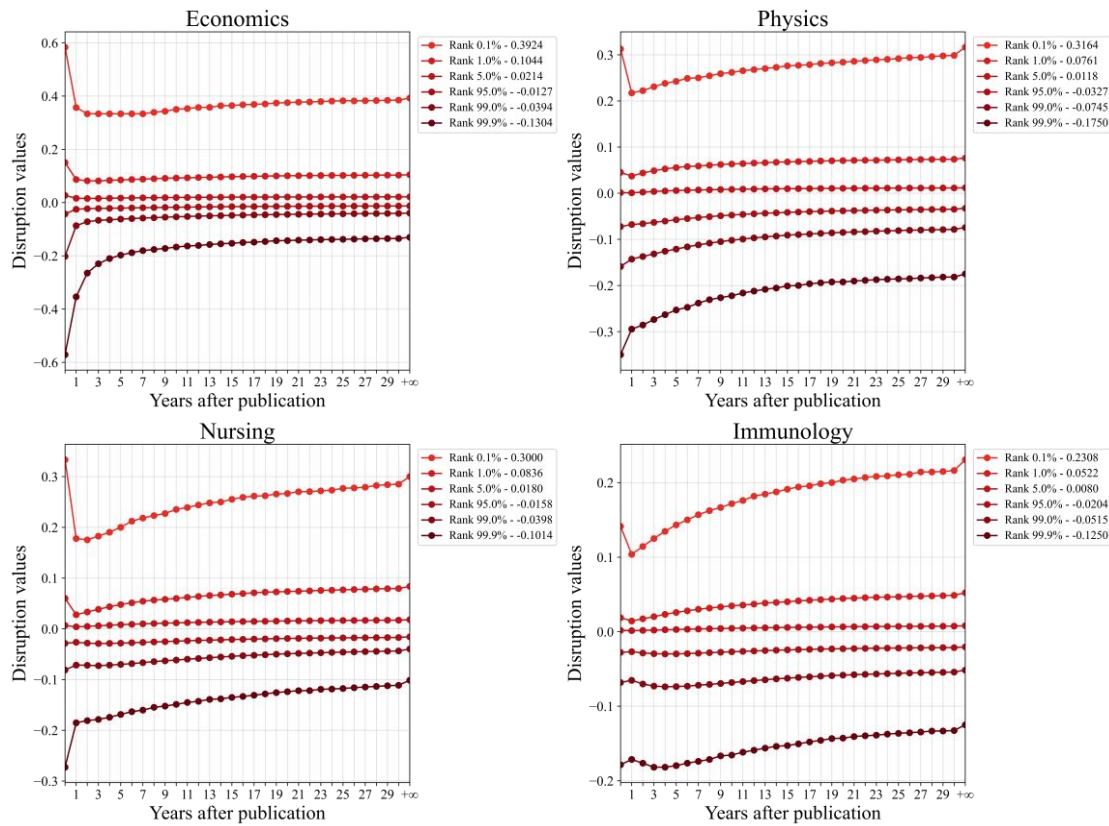

Figure 7. Changes in D values across D ranks. The legend includes the corresponding final D values. For visual clarity, we omit intermediate quantiles (90th, 80th, 50th, 20th, and 10th percentiles) as they exhibit similar trends between the 5th and 95th percentiles, which effectively bound the observed value ranges.

Figure 8 presents three key metrics: a baseline consistency ratio exceeding 80%, alongside two upward-convex curves depicting the classification accuracy and data availability ratios across varying citation windows. As illustrated by the yearly consistent ratio, over 80% of publications identified each year—including those in the zeroth year—retain the same classification as their final D values. However, the line representing the available disruptiveness data ratio underscores a significant limitation of shorter citation windows: the D value cannot be calculated for many publications due to insufficient citation data. For example, in economics, only 60% of publications can be evaluated with a 3-year citation window, while a 5-year window includes nearly 80%, and a 10-year window covers more than 90% of publications.

The consistent ratio further demonstrates that with a 3-year window, 50% of publications are correctly classified and maintain the same D sign as their final classification. This proportion increases to 70% with a 5-year window and surpasses 80% with a 10-year window.

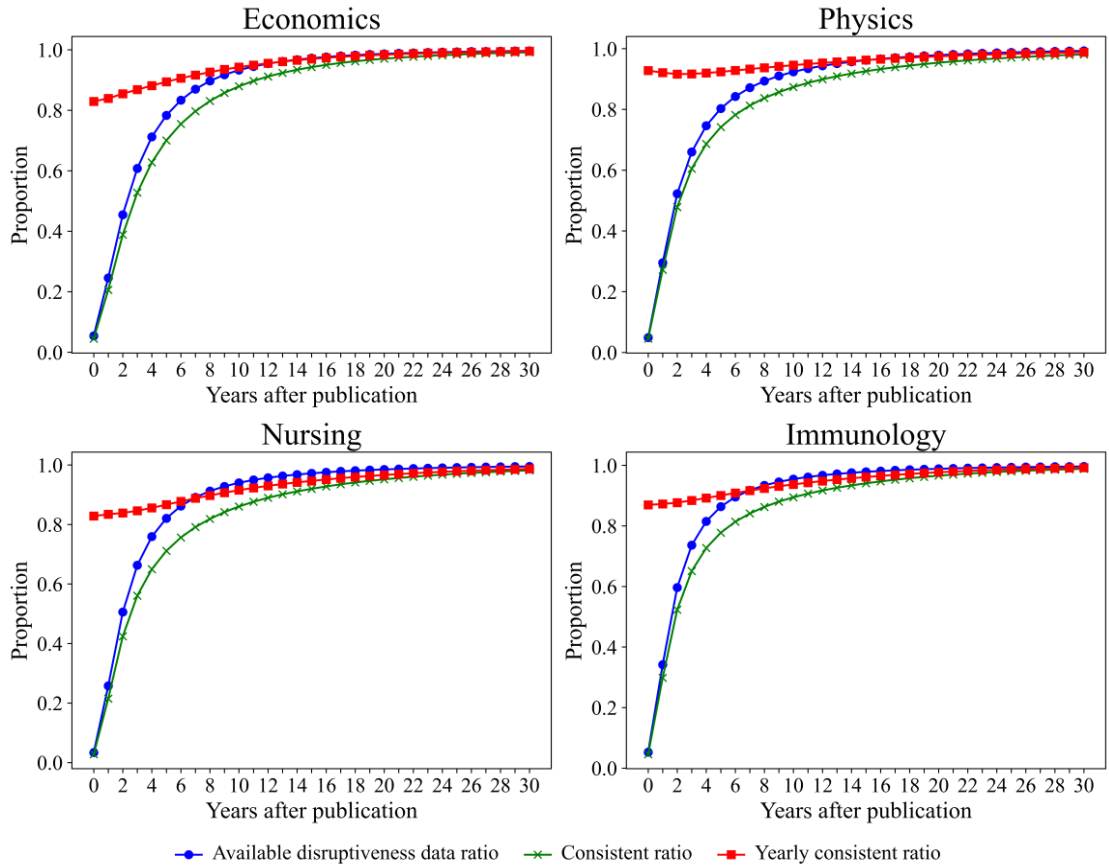

Figure 8. Temporal consistency in term of the sign of D values.

## Stability of D values

It can be observed in Figure 9 that as the constraints on the number of cited references and citations increase, the D value stability time of the publications gradually decreases. For example, in economics, with a threshold of 5, the average stability time of the publications is 5.14 years, which indicates that a 3-year citation window might not be sufficient. Figure 10 reveals that a 3-year citation window can only contain <50% publications, a 10-year citation window can include most of the publications, ensuring the completeness of the study. All lines in Figure 10 are close to each other compared with Figure 3. Figure 10 demonstrates that the stable year is similar across all groups, and it is impractical to include all publications since the stable year exhibits an exponential relationship with the inclusion percentage, with a sharp increase occurring at around 80%. We see from Figure 1 that only considering those with ≥5 references and citations would cause a large exclusion of publications for further D calculations. Hence, the optimal solution is NOT to set any threshold for the citation window to include as many publications as possible, which subsequently leads to incomparability among publications. As a comprise, we argue that one should set a large threshold (e.g., ten) for the citation window to diminish such loss (caused by the time window) regarding the number of publications.

As shown in Figure 9, the D value stability time of publications is influenced differently by the number of cited references and citations. Specifically, as the number of citations increases, the stability time tends to rise, while an increase in the number of cited references leads to a reduction in stability time.

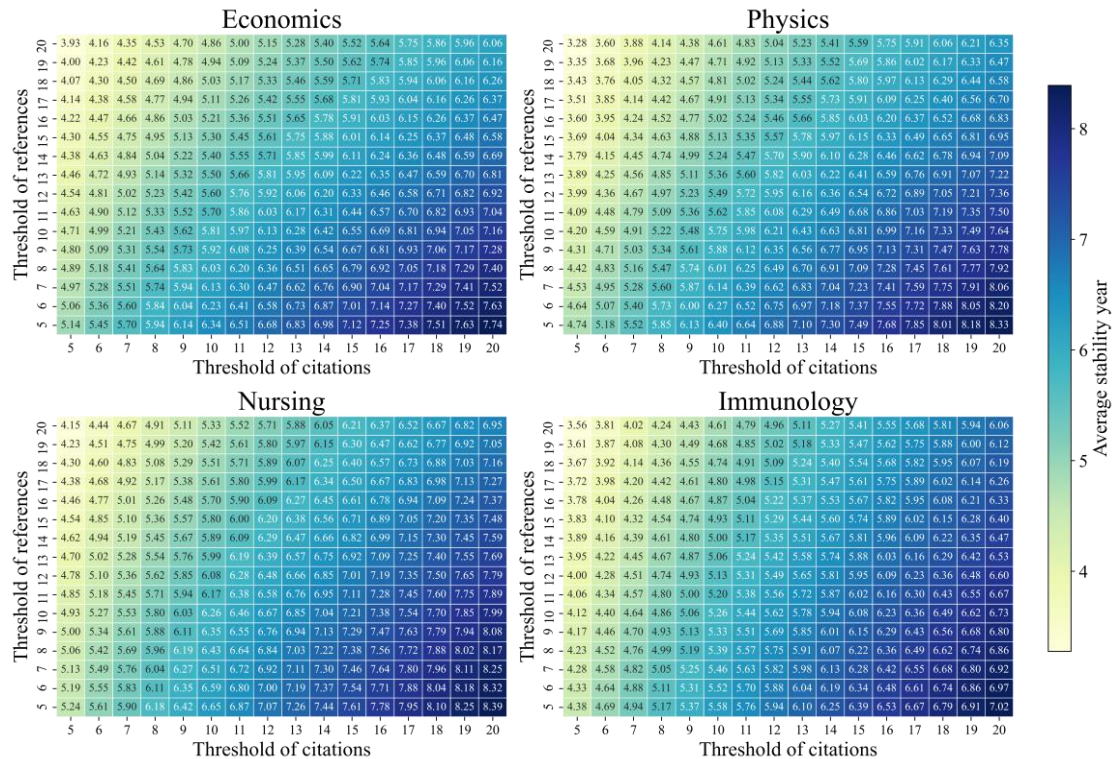

Figure 9. The average D stability time of publications selected under different thresholds.

Figure 10 further illustrates that a 3-year citation window includes no more than half of the publications, while a 10-year citation window covers the majority, ensuring the completeness of the study. Compared to Figure 3 and Figure 10, the lines in Figure 10 are closely aligned, indicating that the stable year is similar across the groups with different thresholds of cited references and citations. However, it is impractical to include all publications, as the stable year exhibits a fast-increasing relationship with the inclusion percentage, with a sharp increase occurring around the 80% threshold. In summary, since D values should not be calculated for most publications, setting a larger threshold (e.g., of ten) for the citation window to maximize inclusion is an effective approach.

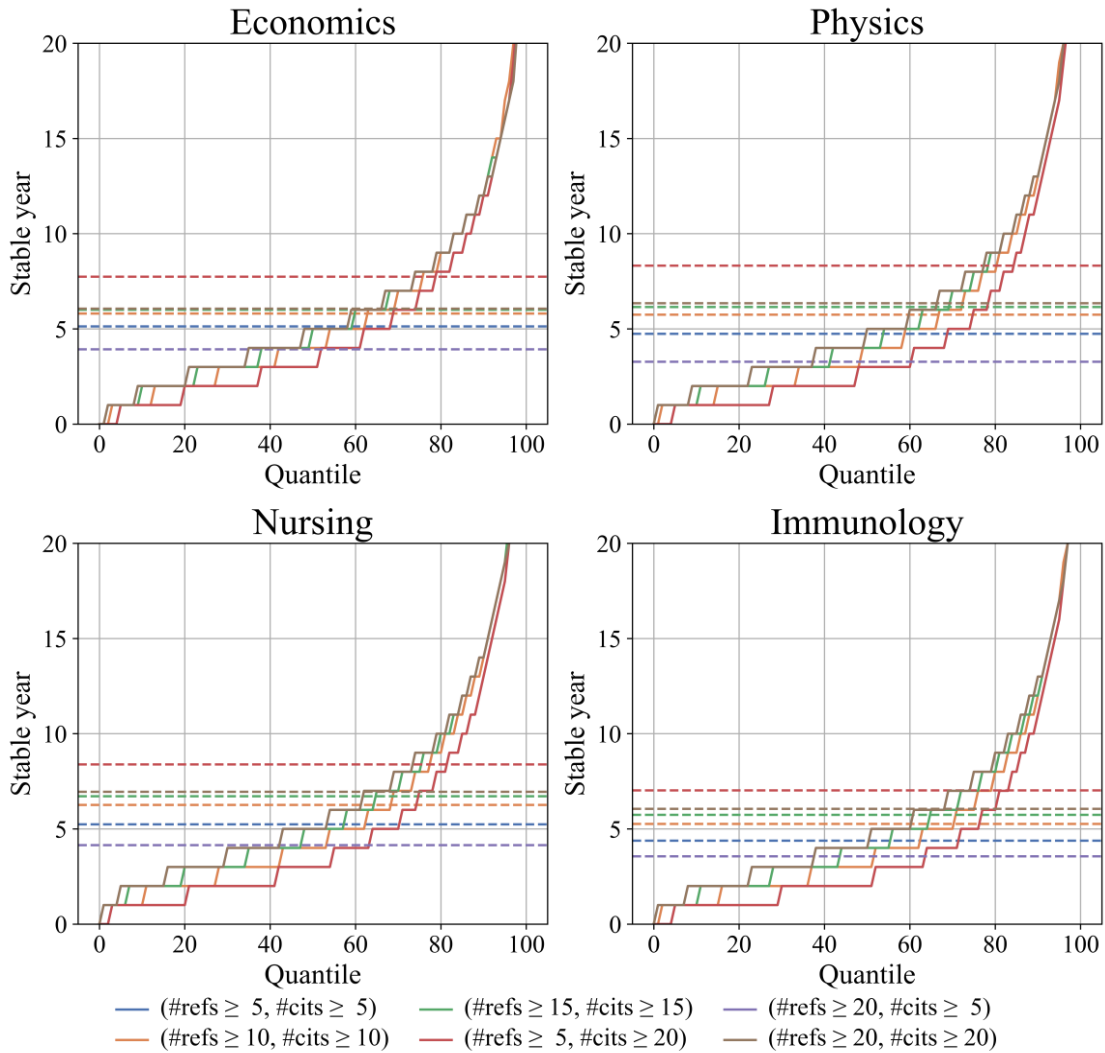

Figure 10. Quantile analysis for the stability time of D. The figure presents the time length (in years) required for publications to stabilize their D values, focusing on selected combinations of reference and citation thresholds. The x-axis represents the quantiles, and the y-axis represents the corresponding time values. The horizontal line represents the average time (the same as Figure 9) of its corresponding color combination.

## Change of D values

The stability discussed above pertains solely to the consistency of the D value signs (above or below zero). Here, we examine the degree of change in D values, as illustrated in Figure 11, which demonstrates a rapid decline in the degree of change over time, with the curve exhibiting a pronounced downward convexity. The percentage change in D values remains above 20% between the fifth and sixth years, while it decreases to approximately 10% between the tenth and eleventh years. Additionally, the figure reveals the absolute magnitude of changes in D values. By the fifth year, the change exceeds 0.001, indicating significant fluctuations in the early stages.

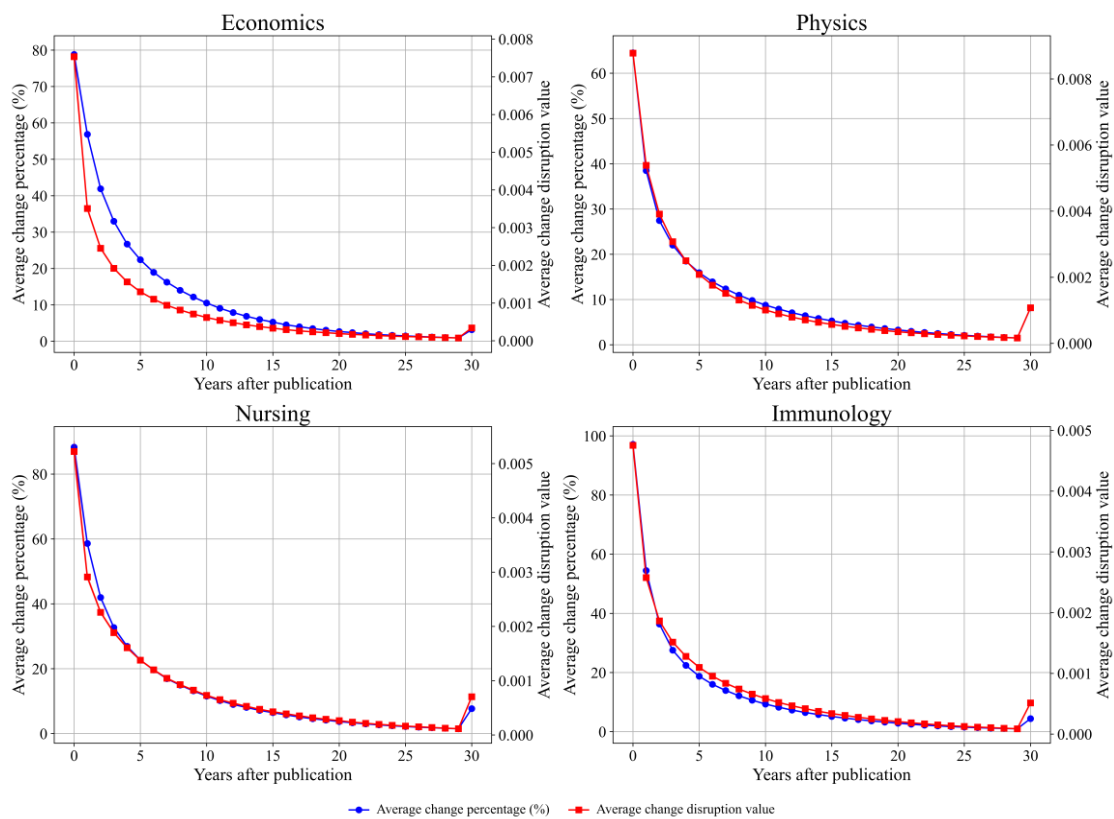

Figure 11. The average change of D values across the years after publishing.

In practice, disruption values of publications across different disciplines may not be directly comparable. As a result, researchers proposed relative rankings of publications' disruptive and consolidating nature (Yu et al., 2024). To address this issue, we normalize the D values of publications by year, as described in the Methods section. Figure 12 illustrates the average change in the normalized rankings of publications, along with the average changes for publications that either rise or fall in rankings each year. The results reveal that publications with declining rankings exhibit smaller changes in early years, while those with rising rankings show larger initial changes (approximately 13% in economics). However, these changes rapidly converge to 4% or even 2% over time. In later years, publications with falling rankings experience larger changes compared to those with rising rankings. Initially, the overall change line lies between the lines for rising and falling rankings. However, after the third year, it stabilizes at the lower end, suggesting that a significant number of publications no longer undergoes changes in their rankings.

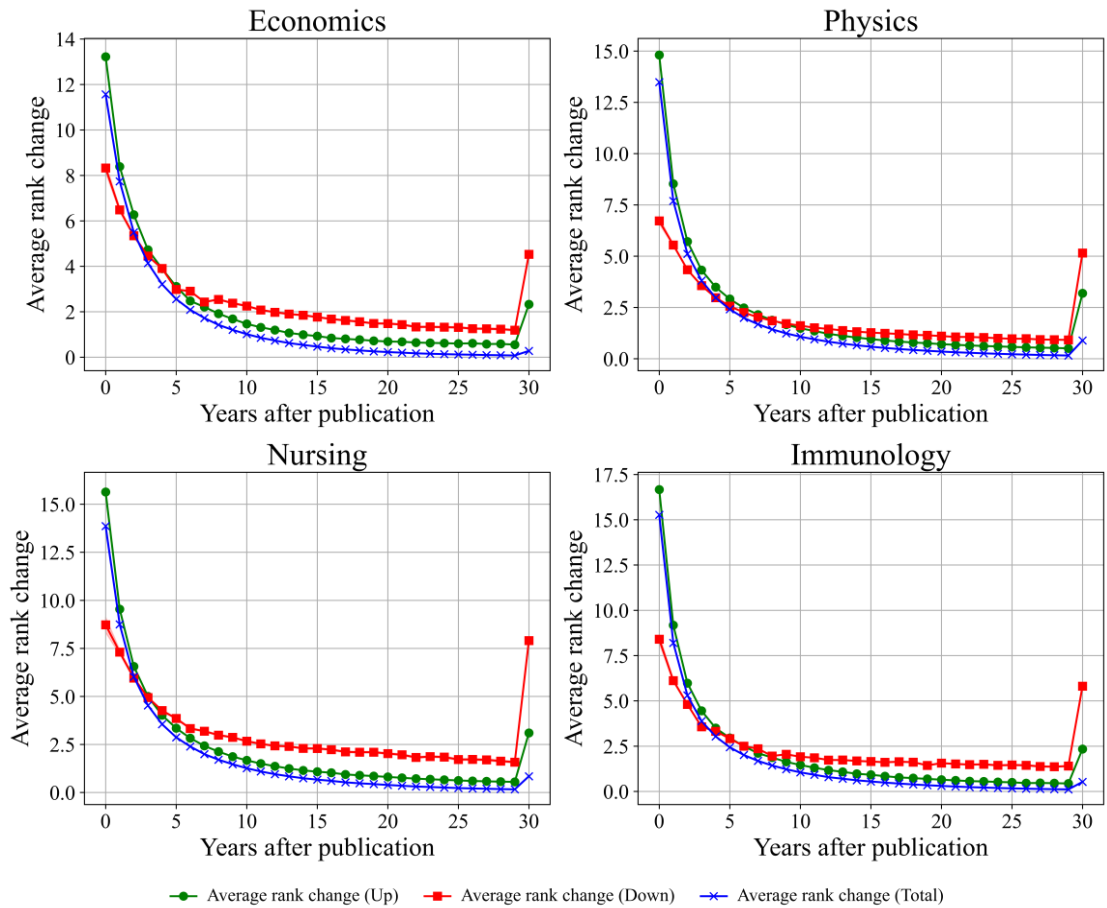

Figure 12. Average change of normalized ranking across years after publishing.

Based on the final D value rankings, Figure S2 reveals that the most disruptive publications (top >99th percentile) and the most consolidating publications (bottom <1st percentile) exhibit relatively smaller changes compared to others. Following these are the consolidating publications (top <5th percentile) and disruptive publications (top >95th percentile). This pattern suggests that highly disruptive and highly consolidating publications can be identified earlier, as their changes are more stable. Overall, after 10 years, the change in D values typically decreases to less than 2%. Figure S3 illustrates the disruption ranking changes of highly disruptive publications (final disruptive values in the top 10%) published in different years, while Figure S4 provides a comparison by showing the changes for all publications. In general, younger publications (2010–2015 and 2015–2020) tend to stabilize more rapidly compared to earlier publications. Focusing on older publications, Figure S5 offers a zoomed-in view of Figure S3, highlighting that these publications follow a more linear stabilization trend. The two lines representing early publications (2000–2005 and 2005–2010) are positioned lower by the seventh year, indicating a slower rate of change in economics.

Table 1 reveals that over 80% of publications remain stable, while more than 94% stabilize after a single transition. Fewer than 3% of publications require more than two transitions, with the maximum number of transitions observed being 17. When the analysis is restricted to a 10-year timeframe instead of 30 years, Table S3 demonstrates even greater robustness in the results.

We further examine the scenario where only highly disruptive publications (top 10%) are classified as "disruptive," while all others are labeled as "consolidating." As shown in Table S4, the results under this condition exhibit stronger robustness compared to the $D > 0$ criterion. As illustrated in Table S5, the results for highly consolidating publications (with only the bottom 10% classified as consolidating) also display significant robustness.

Table 1. Distribution of publications in each type.

| | Economics | | Physics | | Nursing | | Immunology | |
|---|---|---|---|---|---|---|---|---|
| | Count | Proportion (%) | Count | Proportion (%) | Count | Proportion (%) | Count | Proportion (%) |
| Category | | | | | | | | |
| Stable | 775041 | 82.55 | 2732778 | 89.52 | 486021 | 80.17 | 1242907 | 84.58 |
| >1_consolidating_disruptive | 55718 | 5.93 | 122180 | 4.00 | 47922 | 7.90 | 85332 | 5.81 |
| >1_disruptive_consolidating | 17273 | 1.84 | 36552 | 1.20 | 9606 | 1.58 | 21093 | 1.44 |
| Consolidating_disruptive | 17777 | 1.89 | 21717 | 0.71 | 10019 | 1.65 | 17155 | 1.17 |
| Disruptive_consolidating | 17699 | 1.89 | 30020 | 0.98 | 11503 | 1.90 | 26516 | 1.80 |
| Highly_unstable | 55322 | 5.89 | 109405 | 3.58 | 41197 | 6.80 | 76554 | 5.21 |
| Transition Count | | | | | | | | |
| 2 | 32285 | 3.44 | 60216 | 1.97 | 22650 | 3.74 | 42485 | 2.89 |
| 3 | 14687 | 1.56 | 31208 | 1.02 | 11978 | 1.98 | 21882 | 1.49 |
| 4 | 5018 | 0.53 | 10117 | 0.33 | 3738 | 0.62 | 7020 | 0.48 |
| 5 | 2153 | 0.23 | 5049 | 0.17 | 1858 | 0.31 | 3439 | 0.23 |
| 6 | 685 | 0.07 | 1613 | 0.05 | 607 | 0.10 | 1032 | 0.07 |
| 7 | 331 | 0.04 | 790 | 0.03 | 250 | 0.04 | 472 | 0.03 |
| 8 | 105 | 0.01 | 245 | 0.01 | 79 | 0.01 | 144 | 0.01 |
| 9 | 35 | 0.00 | 111 | 0.00 | 23 | 0.00 | 58 | 0.00 |
| 10 | 17 | 0.00 | 36 | 0.00 | 7 | 0.00 | 17 | 0.00 |
| 11 | 4 | 0.00 | 13 | 0.00 | 5 | 0.00 | 4 | 0.00 |
| 12 | 0 | 0.00 | 6 | 0.00 | 0 | 0.00 | 0 | 0.00 |
| 13 | 2 | 0.00 | 1 | 0.00 | 1 | 0.00 | 1 | 0.00 |

| 17 | 0 | 0.00 | 0 | 0.00 | 1 | 0.00 | 0 | 0.00 |

# Conclusion

This study systematically investigates the temporal robustness of D across multiple disciplines, focusing on the interplay between citation accumulation patterns, reference thresholds, and the stability of D over varying time windows. The analysis reveals several critical insights. First, the reliability of D is strongly contingent on the length of the citation window. Shorter windows (e.g., 3 years) exhibit significant instability, with low correlation coefficients (<0.8) compared to final D values, while longer windows (e.g., 10 years) demonstrate robust alignments, capturing over 80% of publications that are ultimately classified as highly disruptive. Second, publications with more cited reference counts achieve citations earlier, reducing the time required for D to stabilize. For instance, economics papers with $\geqslant$30 cited references received their first citation nearly a year sooner than those with minimal references, a trend magnified for higher citation thresholds (e.g., a 2.95-year difference for the 30th citation). Third, early identification of highly disruptive or consolidating publications is feasible: the most extreme cases (top/bottom 1–5%) stabilize within 5 years, with 60–80% of highly disruptive papers in economics that are correctly classified by the 10th year. Finally, the threshold for identifying disruption or consolidation evolves temporally, particularly in fields like economics, where early volatility gives way to stability after 5 years, reflecting a dynamic balance between citation maturation and conceptual integration.

The findings of our study carry significant implications for both research evaluation and science policy. The dependency of D on citation windows and the tendency in research evaluation to focus on recent years underscore the risk of relying on short-term data for assessing scientific performance (Wang, 2013). For instance, while 3-year windows may suffice for preliminary screening, they fail to capture the majority of publications' long-term trajectories, risking misclassification. This aligns with prior critiques of premature evaluation in research assessment frameworks (Waltman, 2016). Conversely, the feasibility of early identification for extreme cases (highly disruptive or consolidating works) suggests that targeted interventions—such as funding allocation or translational support—could prioritize papers showing early disruptive or consolidating signals, albeit with caution to avoid overreliance on noisy early data.

The observed disciplinary variations, particularly in economics' fluctuating D value thresholds, highlight the need for field-specific calibration of D. The "rising entry barrier" for disruptiveness and consolidation over time—where longer windows demand higher D values—may reflect cumulative advantage dynamics (Barabási, 2012; Kozlowski et al., 2024). These dynamics are characterized by early-cited papers that attract disproportionate attention later on. The dynamics align with Matthew effect theories but complicate cross-temporal comparisons. The inverse relationship between cited reference counts and stabilization time suggests that well-connected papers (with more cited references) integrate faster into the scientific discourse, potentially due to their alignment with existing paradigms. This raises questions about whether metrics measuring disruption and consolidation inadvertently penalize interdisciplinary or niche work that takes longer to gain traction.

The 10-year citation window emerges as a pragmatic compromise between completeness and timeliness, covering >90% of publications while mitigating early volatility. However, the trade-off between inclusivity (lower citation/reference thresholds) and metric stability warrants careful consideration. For example, stringent thresholds (e.g., ⩾30 citations) reduce stabilization time but exclude younger or less-cited papers, biasing analyses toward established domains. Policymakers and scientometricians have the task of balancing these factors when designing assessment protocols.

While this study advances our understanding of temporal dynamics in D, several limitations merit attention. (1) Future research could explore hybrid models that integrate D with alternative indicators—e.g., textual novelty (Shibayama et al., 2021)—to compensate for citation-based biases. (2) Longitudinal case studies of specific disruptive or consolidating papers could elucidate possible mechanisms driving early stabilization, such as institutional endorsement or media coverage. The development of dynamic, time-adjusted indices measuring disruption and consolidation, which account for evolving citation baselines, could enhance cross-year comparability. (3) Expanding the scope to non-English publications and non-journal outputs (preprints, patents) would test the generalizability of our findings and may better reflect global scientific practices.

# Acknowledgments

Hongkan Chen and Yi Bu are supported by the National Natural Science Foundation of China (#72104007, #72474009, and #72174016). Hongkan Chen is also supported by the National Social Science Foundation of China (#24ZDA078). The authors declare no competing interests.

# Supplementary Information

Table S1. Descriptive statistics for the used dataset.

| Field | Total number of papers | Publication year range | Publication year span | Mean year | Median year |
|---|---|---|---|---|---|
| Economics | 7748163 | 1019-2025 | 1007 | 2006.1 | 2011 |
| Physics | 9961143 | 1500-2025 | 526 | 2000.6 | 2006 |
| Nursing | 1961284 | 208-2025 | 1818 | 2002.6 | 2009 |
| Immunology | 3825014 | 208-2025 | 1818 | 2003.5 | 2009 |

| Field | Mean citations | Median citations | Minimum citations | Maximum citations |
|---|---|---|---|---|
| Economics | 7.4 | 0 | 0 | 59212 |
| Physics | 14.9 | 1 | 0 | 154914 |
| Nursing | 15.9 | 1 | 0 | 42755 |
| Immunology | 21.8 | 3 | 0 | 22949 |

| Field | Mean number of cited references | Median number of cited references | Minimum number of cited references | Maximum number of cited references |
|---|---|---|---|---|
| Economics | 6.8 | 0 | 0 | 8697 |
| Physics | 13.3 | 5 | 0 | 3706 |
| Nursing | 15.1 | 2 | 0 | 2694 |
| Immunology | 20.3 | 8 | 0 | 5929 |

Table S2. Average time to achieve minimum citation counts across cited reference and citation thresholds in economics. The table presents the distribution of publications across different achieved years, stratified by varying combinations of minimum cited references and citations.

| Year | Combination of minimum (cited references, citations) | | | | | | | |
|---|---|---|---|---|---|---|---|---|
| | (1, 1) | (5, 5) | (10, 10) | (15, 15) | (20, 20) | (1, 30) | (30, 1) | (30, 30) |
| 0 | 533027 | 50896 | 14587 | 6794 | 3644 | 2109 | 192193 | 1152 |
| 1 | 554018 | 179788 | 67480 | 33653 | 19162 | 12272 | 160504 | 7498 |
| 2 | 280533 | 196115 | 99349 | 55331 | 33418 | 24482 | 55982 | 13801 |
| 3 | 144005 | 143702 | 92614 | 58314 | 36775 | 31177 | 0 | 15830 |
| 4 | 0 | 97980 | 72399 | 50099 | 33617 | 33381 | 0 | 14990 |
| 5 | 0 | 66760 | 53189 | 39195 | 27535 | 32108 | 0 | 12792 |
| 6 | 0 | 47000 | 39705 | 29574 | 21359 | 28684 | 0 | 10028 |
| 7 | 0 | 0 | 28973 | 22458 | 16192 | 25068 | 0 | 7814 |
| 8 | 0 | 0 | 0 | 16831 | 12575 | 21480 | 0 | 6091 |
| 9 | 0 | 0 | 0 | 0 | 0 | 18206 | 0 | 0 |
| 10 | 0 | 0 | 0 | 0 | 0 | 15276 | 0 | 0 |
| 11 | 0 | 0 | 0 | 0 | 0 | 12761 | 0 | 0 |

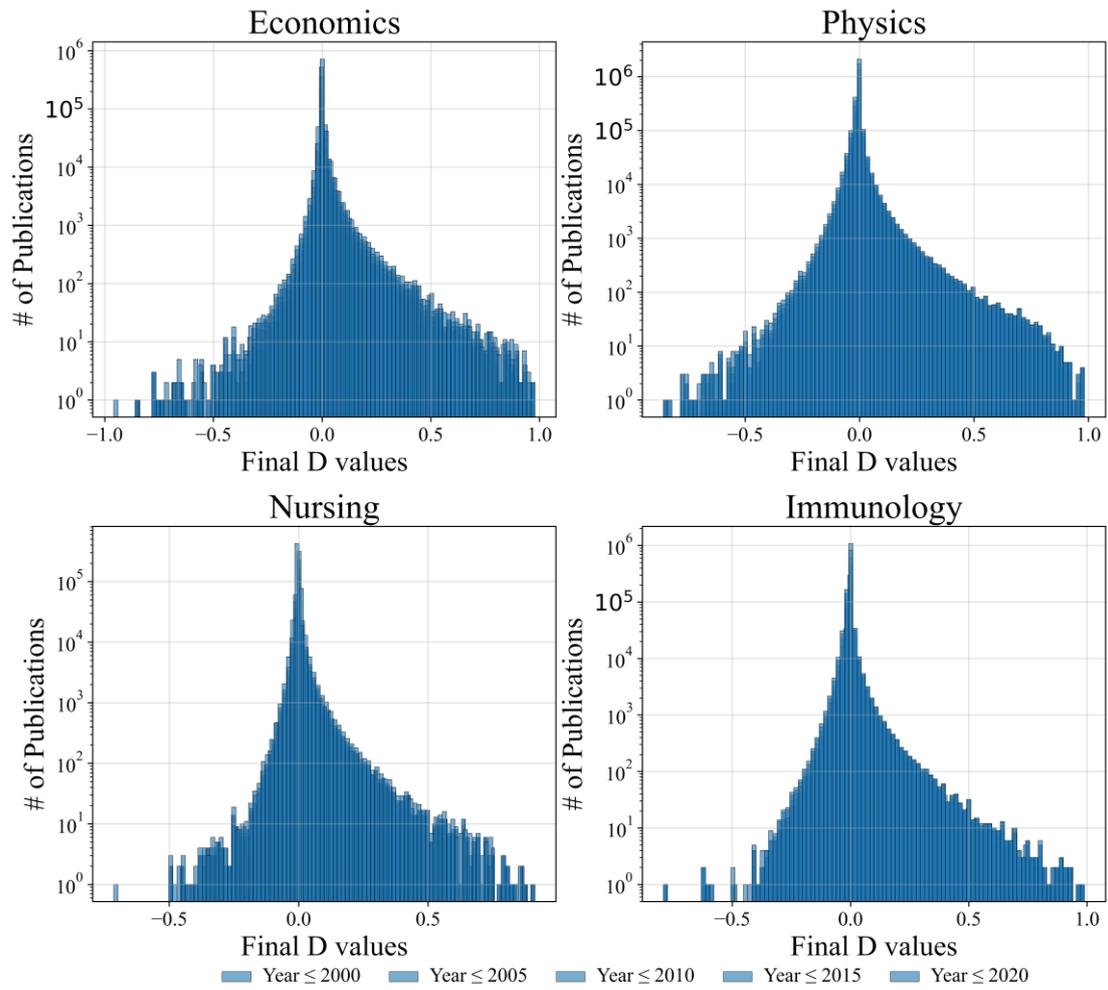

Figure S1. Field-specific distributions of final D values across publications from different years.

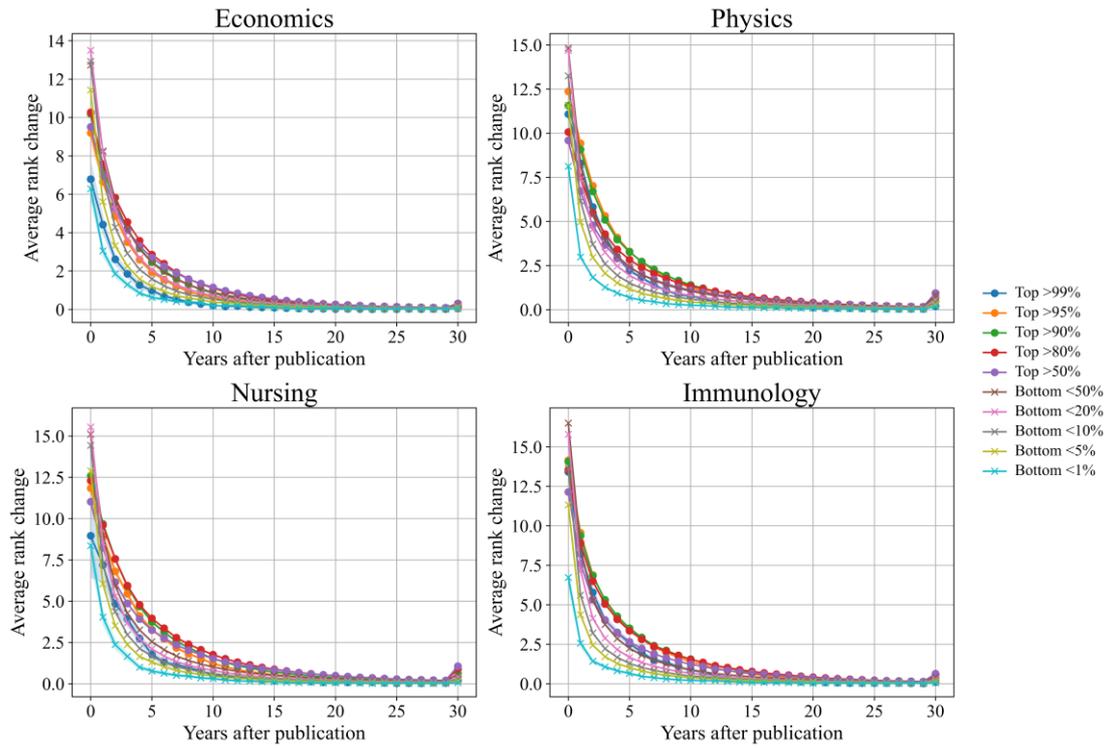

Figure S2. Change of D values across different D-level groups.

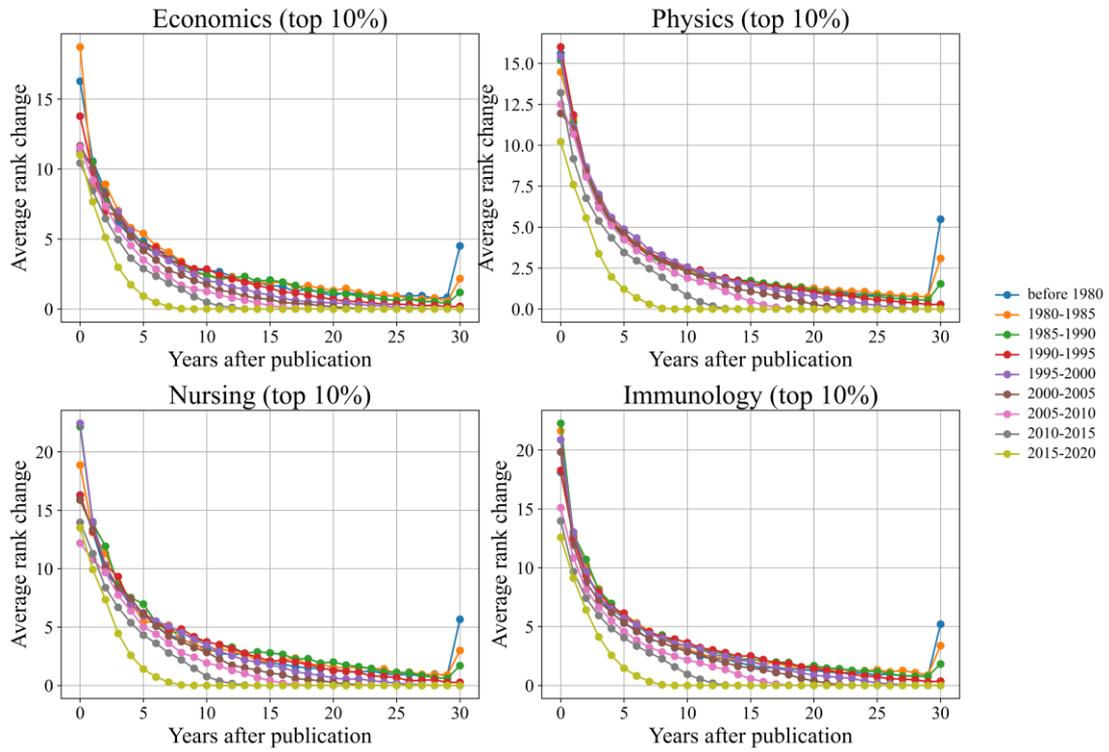

Figure S3. Change of D for articles with D values within the top 10% papers published in different publication years.

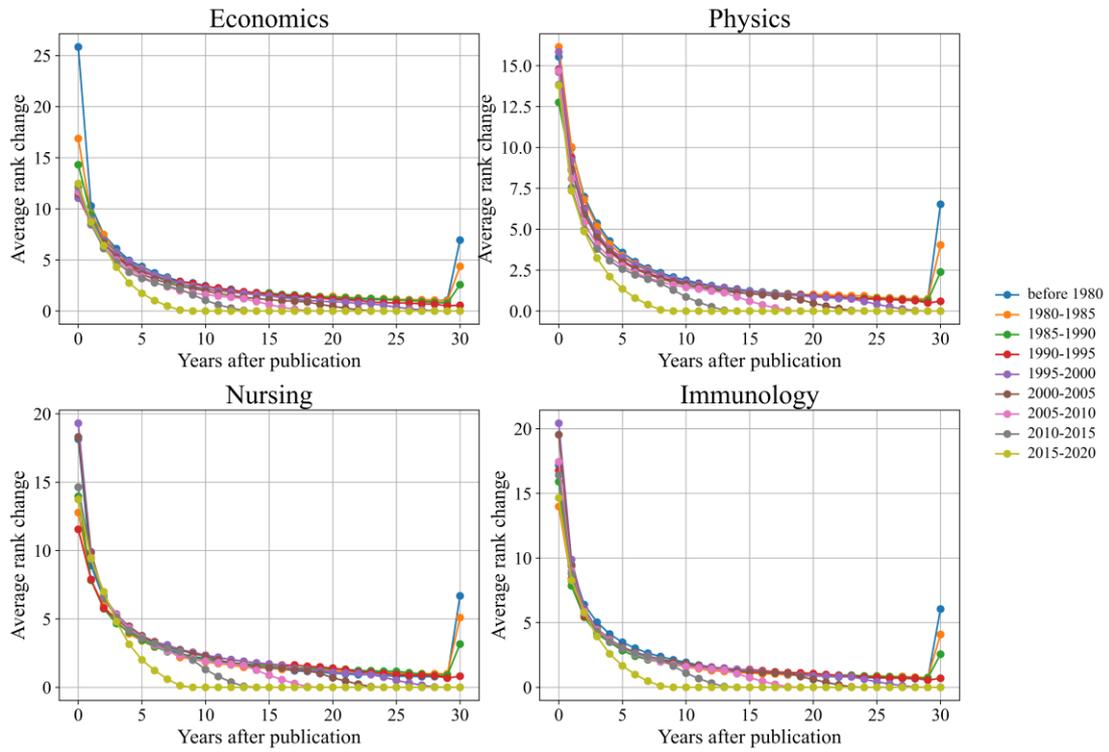

Figure S4. Change of D for articles published in different publication years.

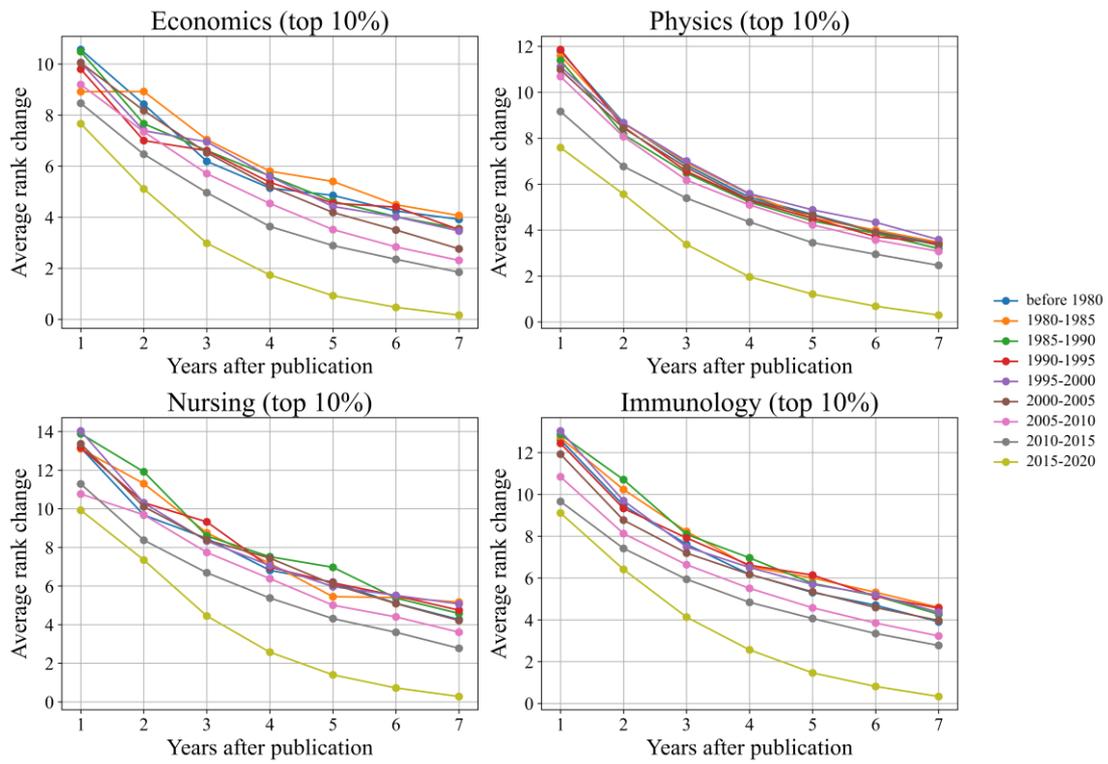

Figure S5. Zoom shots of Figure S3.

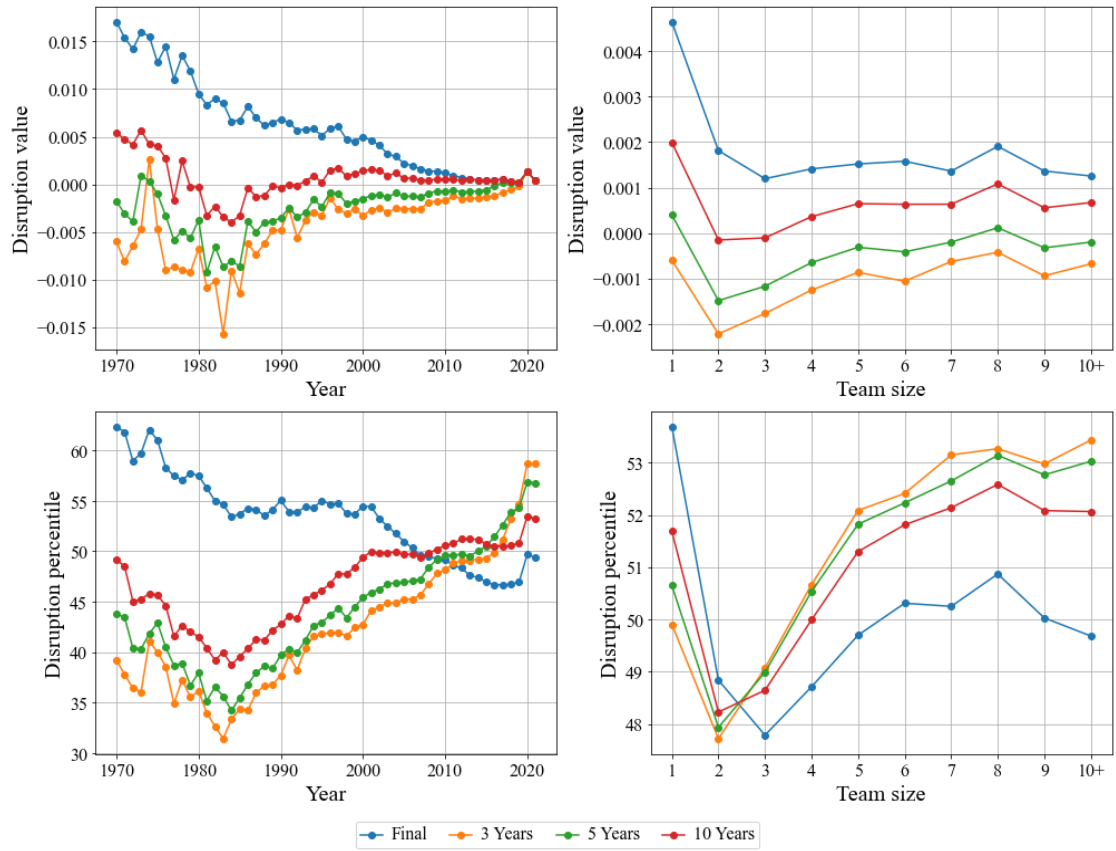

Figure S6. These figures follow Wu et al. (2019) and Park et al. (2023) to explore D values across different years and team size.

Table S3. Distribution of publications across various types

(D values are based on a 10-year citation window).

|  | Economics | | Physics | | Nursing | | Immunology | |
| --- | --- | --- | --- | --- | --- | --- | --- | --- |
|  | Count | Proportion (%) | Count | Proportion (%) | Count | Proportion (%) | Count | Proportion (%) |
| Category | | | | | | | | |
| Stable | 812130 | 86.50 | 2842591 | 93.12 | 520607 | 85.87 | 1304539 | 88.77 |
| >1_consolidating_disruptive | 36296 | 3.87 | 65883 | 2.16 | 26315 | 4.34 | 47216 | 3.21 |
| >1_disruptive_consolidating | 16358 | 1.74 | 29553 | 0.97 | 9979 | 1.65 | 20897 | 1.42 |
| Consolidating_disruptive | 17901 | 1.91 | 22166 | 0.73 | 10079 | 1.66 | 17401 | 1.18 |
| Disruptive_consolidating | 19186 | 2.04 | 32507 | 1.06 | 13000 | 2.14 | 29204 | 1.99 |
| Highly_unstable | 36959 | 3.94 | 59952 | 1.96 | 26288 | 4.34 | 50300 | 3.42 |
| Transition Count | | | | | | | | |
| 2 | 25764 | 2.74 | 41377 | 1.36 | 17982 | 2.97 | 34141 | 2.32 |
| 3 | 8450 | 0.90 | 14153 | 0.46 | 6242 | 1.03 | 11981 | 0.82 |
| 4 | 2132 | 0.23 | 3435 | 0.11 | 1577 | 0.26 | 3241 | 0.22 |
| 5 | 521 | 0.06 | 836 | 0.03 | 395 | 0.07 | 767 | 0.05 |
| 6 | 79 | 0.01 | 131 | 0.00 | 82 | 0.01 | 145 | 0.01 |
| 7 | 11 | 0.00 | 17 | 0.00 | 9 | 0.00 | 20 | 0.00 |
| 8 | 2 | 0.00 | 3 | 0.00 | 1 | 0.00 | 5 | 0.00 |

Table S4. Distribution of publications across various types

(only top 10% D values are considered as disruptive).

|  | Economics | | Physics | | Nursing | | Immunology | |
|---|---|---|---|---|---|---|---|---|
|  | Count | Proportion (%) | Count | Proportion (%) | Count | Proportion (%) | Count | Proportion (%) |
| Category | | | | | | | | |
| Stable | 873266 | 93.02 | 2696281 | 88.33 | 539896 | 89.05 | 1284505 | 87.41 |
| >1_consolidating_disruptive | 16662 | 1.77 | 51916 | 1.70 | 10540 | 1.74 | 28038 | 1.91 |
| >1_disruptive_consolidating | 11607 | 1.24 | 97930 | 3.21 | 16315 | 2.69 | 46023 | 3.13 |
| Consolidating_disruptive | 7744 | 0.82 | 14335 | 0.47 | 3732 | 0.62 | 8877 | 0.60 |
| Disruptive_consolidating | 6337 | 0.67 | 83304 | 2.73 | 12524 | 2.07 | 39610 | 2.70 |
| Highly_unstable | 23214 | 2.47 | 108886 | 3.57 | 23261 | 3.84 | 62504 | 4.25 |
| Transition Count | | | | | | | | |
| 2 | 13394 | 1.43 | 60187 | 1.97 | 12482 | 2.06 | 34963 | 2.38 |
| 3 | 5978 | 0.64 | 29035 | 0.95 | 6333 | 1.04 | 16312 | 1.11 |
| 4 | 2252 | 0.24 | 10976 | 0.36 | 2378 | 0.39 | 6355 | 0.43 |
| 5 | 944 | 0.10 | 5066 | 0.17 | 1178 | 0.19 | 2873 | 0.20 |
| 6 | 388 | 0.04 | 2079 | 0.07 | 507 | 0.08 | 1166 | 0.08 |
| 7 | 149 | 0.02 | 897 | 0.03 | 229 | 0.04 | 493 | 0.03 |
| 8 | 73 | 0.01 | 401 | 0.01 | 94 | 0.02 | 211 | 0.01 |
| 9 | 21 | 0.00 | 159 | 0.01 | 28 | 0.00 | 78 | 0.01 |
| 10 | 10 | 0.00 | 57 | 0.00 | 17 | 0.00 | 32 | 0.00 |
| 11 | 4 | 0.00 | 14 | 0.00 | 9 | 0.00 | 12 | 0.00 |
| 12 | 1 | 0.00 | 10 | 0.00 | 4 | 0.00 | 7 | 0.00 |
| 13 | 0 | 0.00 | 3 | 0.00 | 1 | 0.00 | 1 | 0.00 |

| 14 | 0 | 0.00 | 1 | 0.00 | 0 | 0.00 | 1 | 0.00 |
| 15 | 0 | 0.00 | 1 | 0.00 | 1 | 0.00 | 0 | 0.00 |

Table S5. Distribution of publications across various types

(only bottom 10% D values are considered as consolidating).

|  | Economics | | Physics | | Nursing | | Immunology | |
|---|---|---|---|---|---|---|---|---|
|  | Count | Proportion (%) | Count | Proportion (%) | Count | Proportion (%) | Count | Proportion (%) |
| Category | | | | | | | | |
| Stable | 858195 | 91.41 | 2773603 | 90.86 | 547186 | 90.25 | 1345599 | 91.56 |
| >1_disruptive_consolidating | 31137 | 3.32 | 92480 | 3.03 | 18435 | 3.04 | 33785 | 2.30 |
| >1_consolidating_disruptive | 7457 | 0.79 | 30292 | 0.99 | 5955 | 0.98 | 13825 | 0.94 |
| Disruptive_consolidating | 12529 | 1.33 | 39420 | 1.29 | 8620 | 1.42 | 20269 | 1.38 |
| Consolidating_disruptive | 4075 | 0.43 | 12693 | 0.42 | 3048 | 0.50 | 6913 | 0.47 |
| Highly_unstable | 25437 | 2.71 | 104164 | 3.41 | 23024 | 3.80 | 49166 | 3.35 |
| Transition Count | | | | | | | | |
| 2 | 14685 | 1.56 | 55794 | 1.83 | 12066 | 1.99 | 26212 | 1.78 |
| 3 | 6128 | 0.65 | 26057 | 0.85 | 5905 | 0.97 | 12400 | 0.84 |
| 4 | 2767 | 0.29 | 12362 | 0.40 | 2779 | 0.46 | 5726 | 0.39 |
| 5 | 1027 | 0.11 | 5360 | 0.18 | 1242 | 0.20 | 2713 | 0.18 |
| 6 | 530 | 0.06 | 2696 | 0.09 | 605 | 0.10 | 1209 | 0.08 |
| 7 | 169 | 0.02 | 1053 | 0.03 | 247 | 0.04 | 552 | 0.04 |
| 8 | 93 | 0.01 | 528 | 0.02 | 105 | 0.02 | 224 | 0.02 |
| 9 | 28 | 0.00 | 168 | 0.01 | 51 | 0.01 | 79 | 0.01 |
| 10 | 9 | 0.00 | 92 | 0.00 | 13 | 0.00 | 35 | 0.00 |
| 11 | 1 | 0.00 | 34 | 0.00 | 9 | 0.00 | 9 | 0.00 |

| 12 | 0 | 0.00 | 12 | 0.00 | 1 | 0.00 | 4 | 0.00 |
| 13 | 0 | 0.00 | 2 | 0.00 | 1 | 0.00 | 2 | 0.00 |
| 14 | 0 | 0.00 | 4 | 0.00 | 0 | 0.00 | 1 | 0.00 |
| 15 | 0 | 0.00 | 2 | 0.00 | 0 | 0.00 | 0 | 0.00 |